\documentclass[10pt,aps,prl,twocolumn,superscriptaddress]{revtex4}

\usepackage[utf8]{inputenc}
\usepackage[linktocpage=true, colorlinks=true, urlcolor=blue, linkcolor=blue, citecolor=blue]{hyperref}
\usepackage{graphicx}
\usepackage{amsmath,amssymb}
\usepackage{color}
\usepackage{subfigure}
\usepackage{url}

\graphicspath{{Images/}}

\begin{document}

\title{Quantum revival for elastic waves in thin plate}

\author{Marc Dubois}
\author{Gautier Lefebvre}
\affiliation{
Institut Langevin, ESPCI ParisTech CNRS UMR7587, 1 rue Jussieu, 75238 Paris cedex 05, France\\
}
\author{Patrick Sebbah}
\affiliation{
Institut Langevin, ESPCI ParisTech CNRS UMR7587, 1 rue Jussieu, 75238 Paris cedex 05, France\\
}
\affiliation{
Department  of  Physics,  The  Jack  and  Pearl  Resnick  Institute  for
Advanced  Technology,  Bar-Ilan  University,  Ramat-Gan,  5290002  Israel\\
}

\date{\today}

\begin{abstract}
Quantum revival is described as the time-periodic reconstruction of a wave packet initially localized in space and time. This effect is expected in finite-size systems which exhibits commensurable discrete spectrum such as the infinite quantum well. Here, we report on the experimental observation of full and fractional quantum revival for classical waves in a two dimensional cavity. We consider flexural waves propagating in thin plates, as their quadratic dispersion at low frequencies mimics the dispersion relation of quantum systems governed by Schr\"{o}dinger equation. Time-dependent excitation and measurement are performed at ultrasonic frequencies and reveal a periodic reconstruction of the initial elastic wave packet. 
\end{abstract}
\maketitle

In an infinite quantum well or a closed wave cavity, an eigenstate is a stationary oscillating field pattern, which self-reproduces at its eigenfrequency. A set of eigenstates can be adjusted in amplitude and phase to interfere constructively in a small region of space and synthesize a wave packet. Because it is composed of several eigenstates, the wave packet is not a stationary wavefunction that remains confined in space but will spread over the entire system. The time needed by this wave packet to get to the “maximum spreading” is often called the Ehrenfest time. Once this time is reached, a question is raised: can the spreading reverse and the wavefunction fold back to reconstruct the initial wave packet, in the manner of a time reversal operation \cite{Robinett2004}. The answer is positive for dispersive systems, namely for systems with not uniformly distributed discrete energy levels. This fundamental condition for observation of the quantum revival effect is naturally fulfilled in quantum systems following Schr\"{o}dinger equation. For this reason, the ability of quantum systems to fold back to their initial state triggered a lot of interest in the quantum physics community. The quantum revival effect was first predicted in the Jaynes-Cummings model \cite{Jaynes1963,Eberly1980,Narozhny1981}, in Rydberg wave packets \cite{Parker1986,Averbukh1989} and in chaotic quantum systems \cite{Tomsovic1997}. Its first experimental observation was achieved in low dimensional systems, such as Rydberg atoms \cite{Rempe1987,Yeazell1990,Wals1994}, electrons in graphene under magnetic field \cite{Krueckl2009}, molecular systems \cite{Vrakking1996} and cavity quantum electrodynamics \cite{Kirchmair2013}. In optics, the Talbot effect has been identified as a classical spatial analog of the temporal quantum revival effect. In 1836, Henry Fox Talbot reported that the image of a grating is identically reproduced when illuminated by a plane wave \cite{Talbot1836}. The theory of the Talbot effect was later derived in the paraxial limit, where the Helmholtz equation reduces to a parabolic Schr\"{o}dinger-type equation \cite{Winthrop1965,Berry1996}. Dispersion in optical waveguides served also to demonstrate this effect \cite{Efremidis2005,DellaValle2009}. However, not all physical aspects of the quantum revival are captured by the Talbot effect. For example, the reconstruction distance varies with the incident wavelength in the Talbot effect, while the revival time is only a function of the dimensions of the cavity. Two dimensional (2D) quantum revival is also an experimental challenge. Apart from theoretical predictions \cite{Aronstein1997}, there is no experimental demonstration of the revival effect in 2D cavities.

In this article, we report on the experimental observation of elastic wave packet revival in a thin plate. In the limit of wavelengths much larger than the plate thickness, three-dimensional elastic wave equations reduce to the 2D bi-harmonic Kirchhoff-Love plate equation. This equation presents a natural quadratic dispersion relation which mimics the dispersion relation of quantum systems. Full and fractional revival are both observed and discussed. Time domain measurements are supported by simple theory, which allow us to provide with a new understanding of the revival phenomenon for classical waves.

\begin{figure}
\centering
\includegraphics[width=8.5cm]{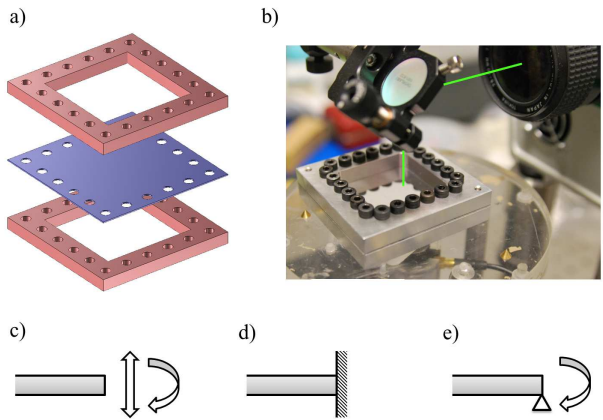}
\caption{a) Exploded view of the elastic cavity. The thin Duraluminium plate is sketched in blue while the two frames which enforce rigid boundaries are represented in red.  b) Picture of the experimental setup. A heterodyne laser interferometer is used to probe the vertical displacement of the plate. The contact transducer is located under the sample and is not visible in this picture. c,d,e) Sketch of the plate boundary conditions, free, rigid, and simply supported. Edge vertical displacement and rotation are depicted by the arrows.}
\label{figure0}
\end{figure}

In the case where the wavelength is large compare to the plate thickness, only two cutoff-free modes can propagate in the plate: the symmetric, $S_0$, and anti-symmetric, $A_0$, Lamb modes. The $S_0$ mode polarization is mostly longitudinal with small off-plane displacement, in contrast to the $A_0$ mode which mainly generates an in-plane movement. In our experiment, the contact transducer produces only vertical displacement in order to excite selectively the $A_0$ mode. In the low frequency regime, the Kirchhoff-Love theory describes the vertical displacement with a bi-harmonic equation that involves a bi-Laplacian spatial operator, $\Delta^2$ \cite{Royer1996}. This fourth order operator leads to a quadratic dispersion relation \eqref{eq_Dispersion}:
\begin{align}\label{eq_Dispersion}
	\omega = ak^2 \; \text{ with, } \; a=\sqrt{\frac{Eh^2}{12(1-\nu^2)\rho}}
\end{align}
which involves the Young’s modulus  $E$, the plate thickness $h$, the mass density $\rho$ and the Poisson ratio $\nu$ of the plate material. 
Because of the fourth order elliptical operator, plate boundary conditions require two physical quantities to be set at the edge. Vertical displacement and rotation around the edge axis are commonly used. From these two quantities one can distinguish two extreme cases: the free boundary condition where vertical displacement and rotation are free at the edge, represented in Fig.~\ref{figure0}c, and the rigid boundary condition where both vertical displacement and rotation are set to zero (Fig.~\ref{figure0}d). Between those two conditions lies the simply supported condition where vertical displacement is forbidden but rotation around the edge is kept as shown in Fig.~\ref{figure0}e. These three representations are conventionally used to depict the boundary conditions. They do not represent real experimental setups. This specific condition is the only one that does not modify the dispersion relation Eq.~\ref{eq_Dispersion}. Both free and rigid boundary conditions induce a slight deviation from the original dispersion relation due to local modification of the material parameters \cite{Leissa1969}. Unfortunately, simply supported boundary conditions are tedious to set experimentally. In our experiment, we chose to set rigid boundaries that allow the smallest deviation from the pure quadratic dispersion relation. Quantum revival effect in the presence of deviation from pure quadratic law or in the case of finite potential square well has already been studied \cite{Robinett2004,Venugopalan1999}. Both studies concludes that the induced error is cumulative in time. Early times fractional revival dynamics still occur but full revival of the initial wave packet might be deteriorated. We will show that we can overcome such a limitation in our experiment by locating the point source at specific positions in order to reduce the revival time.

A short Gaussian pulse with $500$ kHz carrier frequency and $6 \mu $s duration is synthesized by a waveform generator (Agilent 33120A). The ultrasound excitation is applied to the plate via a contact piezoelectric transducer (Panametrics M109). We cut a 40 mm$\times$40 mm square cavity in a Duraluminium plate with thickness $h=0.5$ mm. Rigid boundary conditions are enforced by two square frames firmly bolted on both faces of the plate as shown in Fig~\ref{figure0}a. Out of plane displacement is measured point by point with a laser heterodyne interferometric probe (THALES SH140) (see Fig.~\ref{figure0}b). A 2D scan (80 x 80 points) of the optical probe is performed over the plate with 0.5 mm step to map the temporal and spatial dependence of the elastic field. The short-pulse propagation in the thin plate gives the impulse response of the system, while a spectral filter centered around $\omega$ is used to obtain the propagation of a narrowband gaussian pulse with adjustable carrier frequency $\omega$. This measurement gives access to the time-dependent vertical displacement field $z(t)=A(t)cos(\omega t+\phi(t))$ with $A(t)$ positive at all times. As the amplitude $A(t)$ is slow compare to the carrier period of the pulse, we perform an envelope detection based on Hilbert transform to have direct access to $A(t)$ and $\phi(t)$. 
In order to detect wave packet full and fractional revivals we compute the following function ($C(t)$) at each time step \eqref{eq_IPR}:
\begin{equation}\label{eq_IPR}
	C(t)=\frac{\sum_{\text{plate}}A(t)^2}{[\sum_{\text{plate}}A(t)]^2}
\end{equation}
$C(t)$ is a measure of the spatial confinement of the elastic field within the cavity. It is unity for field concentrated on a single pixel, and equals $\frac{1}{N^2}$ if the field is equally distributed over the entire surface, where ${N^2}$ is the number of pixels contained in the measurement map. The function $C(t)$ is monitored and is used to detect revival events where the wave packet localizes in one or few locations of the plate. 

\begin{figure}
\centering
\includegraphics[width=9cm]{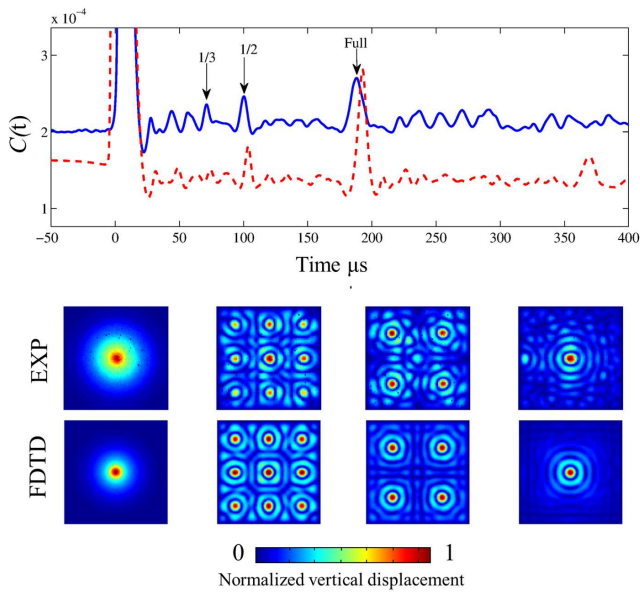}
\caption{\textbf{Top}: Inverse participation ratio of the amplitude of the field as a function of time. Blue solid line: experimental data for the plate. Red dashed line: simulation data for the same plate. Black arrows pinpoint high values of $C(t)$ corresponding to spatial confinement of the field. \textbf{Bottom}: Amplitude snapshots for experiment (upper line) and FDTD simulations (lower line). The first column is the initial pulse, positioned in that case at the center of the square. Other snapshots show amplitude distribution corresponding to fractional revival events. Full time evolution is available in the movie \cite{SuppInf}.}
\label{figure3}
\end{figure}

Figure~\ref{figure3} presents the temporal evolution of $C(t)$ for a point source located at the center of the square plate. The experimental data (Blue solid line) is compared to 3D-finite difference time domain (FDTD) simulations using SimSonic3D software \cite{Virieux1986,Bossy2004} (Red dashed line) for the same plate dimensions and characteristics. Simulation results are obtained on a grid of 100 x 100 points which explains the lower baseline for $C(t)$ (Red dashed line). Regularly spaced peaks in $C(t)$ (pinpointed by black arrows) are identified as fractional and full revival of the wave packet. This is confirmed by the corresponding snapshots of the field amplitude distribution at these specific times.
The first snapshot represents the wave packet at initial time t=0.
Full revival is observed at time $ T_{\text{rev}}$=185$\pm10$  $\mu $s when the wave packet is reformed at the center of the plate (Fig.~\ref{figure3} first right snapshot). The additional rings result from the interference between the converging and diverging waves as the wave packet expands again after $ T_{\text{rev}}$. Actually, the revival is expected to occur periodically. This is confirmed numerically by a second peak observed at $ 2\times T_{\text{rev}}$=370$\pm10$  $\mu $s, which indeed corresponds to a second refocusing of the wavefield at the center of the plate.
Other peaks in $C(t)$ are observed at earlier times. They are identified as fractional revivals as they occur at  fractions of the full revival time. At $ \frac{T_{rev}}{3}$=70$\pm10$  $\mu $s, the third revival is observed, characterized by 9 copies of the source arranged on a square lattice. The quarter revival is detected at $ \frac{T_{rev}}{4}$=100$\pm10$  $\mu $s results in 4 symmetric images of the source. The uncertainty in the time measurements is due to the finite pulse duration. Finally, we verified that the revival time is independent on incident wavelength by reproducing the data analysis for different central frequencies of the wave packet.

\begin{figure}[!h]
\centering
\includegraphics[width=9cm]{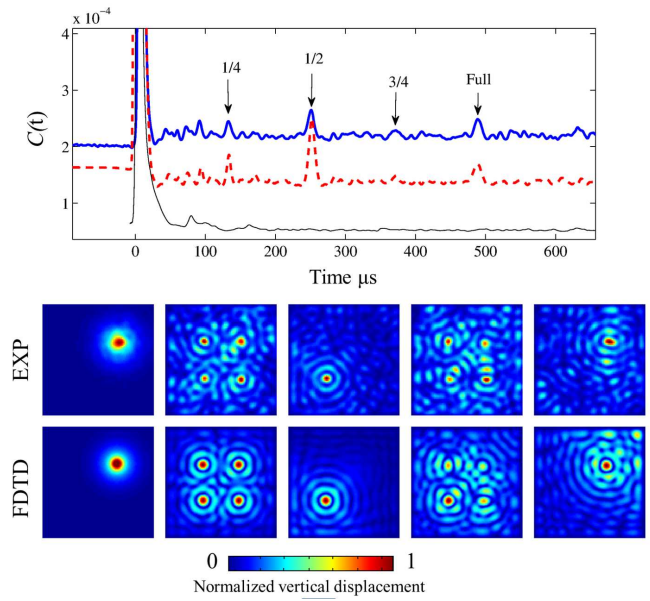}
\caption{Same as \ref{figure2} for a source positioned at one third of the diagonal of the square. Black thin line: $C(t)$ for a membrane (simulations). In that case, dispersion relation is linear and revival does not happen.}
\label{figure2}
\end{figure}

A new experiment is run with a source positioned this time at a third of the diagonal line of the square. Figure~\ref{figure2} presents the temporal evolution of $C(t)$ as well as the field amplitude distribution associated with peaks of $C(t)$. Experimental results and numerical simulations are again compared. Full revival is observed at $ T_{rev}$=480$\pm10$  $\mu $s. This value is different from the revival time found for a source at the center, which indicates that the revival time is dependent on the position of the source. This observation will be explained later.
A four-symmetric-points pattern is observed at time $ \frac{T_{rev}}{4}$=132$\pm10$ $\mu $s. This is again the spatial signature of the quarter revival. At half revival time ($ \frac{T_{rev}}{2}$=250$\pm10$  $\mu $s), the wave packet is reconstructed at the symmetric position of the source with respect to the center of the plate. This fractional revival is not observed in the case of Fig.~\ref{figure3} due to the central position of the source in that case. On the other hand, third revival does not exist in the case of Fig.~\ref{figure2} when the source is at a third of the diagonal. 

When the position of the source is arbitrary, full revival is not observed, while numerical simulations only show very early fractional revivals.

\begin{figure}
\centering
\includegraphics[width=9cm]{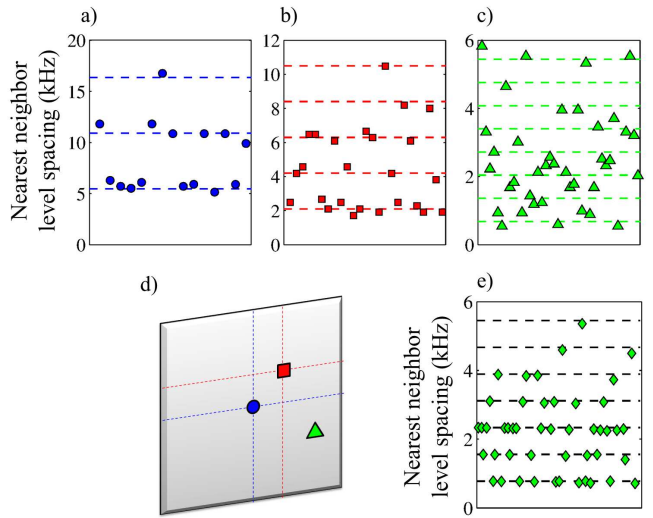}
\caption{Distribution of nearest neighbor level spacing computed from FDTD simulations of the rigid plate with point source positioned a) at the center (0.5$L$,0.5$L$) (from first 15 eigenfrequencies); b) at the third of the diagonal (0.66$L$,0.66$L$) (from first 15 eigenfrequencies); c) at arbitrary position (0.83$L$,0.32$L$) (from first 45 eigenfrequencies). Dashed lines correspond to multiple of the smallest level spacing. d) Sketch of the corresponding source positions. e) Distribution of nearest neighbor level spacing computed from finite element method (FEM) simulations at arbitrary position for a plate with simply supported boundaries (from first 47 eigenfrequencies)}
\label{figure1}
\end{figure}

The general calculation of the revival time for 2D cavities is not simple, compare to 1D systems, since two k-space coordinates are required, associated with two quantum numbers. For the square cavity however, the 2D potential is separable and the cross-terms between the two coordinates disappear. Therefore, the dispersion relation simplifies and the following expression can be deduced for the revival time \cite{Doncheski2003}: 
\begin{align}
\label{eq_Trev}
	T_{rev theo} = \frac{4L^2}{\pi\frac{\delta^2 \omega}{\delta k^2}}.
\end{align}
It can be noticed that this expression is general and does not depend on the initial source position. Doncheski et \textit{al.} \cite{Doncheski2003} have shown that the revival time is reduced when the source seats on a point of symmetry of the square. The revival time at the center of the square is $\frac{T_{\text{rev theo}}}{8}$ and it is $\frac{T_{\text{rev theo}}}{3}$ at the third of the diagonal. 
From Eq.~\ref{eq_Trev} and the characteristics of our plate, $T_{\text{rev theo}}$ =1280 $\mu $s. From the measure of $T_{rev}$ in both cases of Fig.~\ref{figure3} and Fig.~\ref{figure2}, we obtain respectively $T_{\text{rev theo}}$=$ 8\times T_{\text{rev center}}=1480$  $\mu $s and $T_{\text{rev theo}}$=$ 3\times T_{\text{rev diag}}=1440$  $\mu $s
The discrepancy between $ T_{\text{rev theo}}$ measured and the theoretical prediction from Eq.~\ref{eq_Trev} comes from the choice for boundary conditions. Indeed, the separable analytic expression only exist for simply supported boundary conditions \cite{Anselmet2015}. Rigid boundary conditions, as in our experiment, introduce a local deviation of the mechanical parameters near the edges of the plate, which shifts the eigenstates away from the quadratic dispersion and makes the theoretical prediction much more complicated.
\\
\\
A new approach is developed in order to estimate the revival time in our plate with rigid boundary conditions. The dynamics of the wave packet is dictated by the frequency spacing between successive eigenfrequencies. We calculate the first 50 eigenfrequencies of our cavity using 3D-FDTD simulations and compute the distribution of the nearest neighbor level spacing. Figure~\ref{figure1} reveals how they are distributed near multiples of their least common multiple (dashed lines), which is obtained by considering the smallest non zero level spacing. The full revival time is given by the inverse of this smallest level spacing. At this specific time, all the beating between successive eigenfrequencies will oscillate during an integer number of periods, therefore leading to a constructive interference and a full revival of the initial wave packet. This approach confirms that the revival time does not depend on the central frequency of the wave packet but rather on the smallest frequency difference between the eigenvalues excited by the wave packet. We compare, in Fig.~\ref{figure1}a-c, this distribution for three different locations of the point source in our plate.

Figure \ref{figure1}a shows the level spacing distribution obtained numerically with a point source placed at the center of the cavity. In this configuration, the smallest level spacing is $\delta f_{c}$=5.5 kHz, which gives a revival time $ T_{\text{rev center}}$=182 $\mu $s, in excellent agreement with direct experimental measurement in  Fig.\ref{figure3}.

Figure \ref{figure1}b shows the level spacing distribution obtained numerically with a point source located on the third of the square diagonal, as in Fig.~\ref{figure2}. One can see that the smallest frequency spacing is now $\delta f_{d}$=2.06 kHz, corresponding to a larger predicted revival time $ T_{\text{rev diag}}$=484 $\mu $s. This value is also very close to the value obtained experimentally. The level spacings are near multiples of the smallest level spacing $\delta f_{d}$.

Finally, Fig.~\ref{figure1}c shows the level spacing distribution for an arbitrary position of the source. In that case, the distribution is no longer organized around multiple of the smallest level spacing. This means that the commensurability of the eigenfrequencies is lost, and explains why we are not able to observe the full revival for arbitrary initial positions. In order to understand this behavior, we simulate for comparison a plate with simply supported boundaries and compute the level spacing distribution for the same source position. The results, presented in Fig.~\ref{figure1}e, show that the level spacing distribution is commensurable with very little deviation, which ensures the existence of the full revival. The smallest level spacing in that case is $\delta f_{s}$=780 Hz, which gives a full revival at time $ T_{\text{rev s}}$=1282 $\mu $s. This time is very close to the theoretical value calculated with Eq.~\ref{eq_Trev} $T_{\text{rev theo}}$ =1280 $\mu $s.

This analysis offers an interesting explanation of how the location of the initial excitation affects the revival time. We confirm that the shortest revival time is found for a source placed at the center of the cavity and that it increases when the number of symmetries is reduced due to the position of the source. Actually, when the source is positioned at a point of high symmetry, such as the center of the square, the eigenmodes with a node at this position are not excited and therefore do not participate to the building of the wave packet. This results in a new organization of the level spacing distribution with a larger minimum spacing and, consequently, a shorter revival time. This reduction of the revival time was theoretically predicted in \cite{Doncheski2003} and is nicely illustrated here.

To further confirm that nonlinear dispersion is essential for the revival effect, we numerically compare the propagation of an elastic pulse in a plate and in a membrane. In contrast to flexural waves in thin plates, vibrations of membranes are described by an regular Helmholtz equation that leads to a linear dispersion relation. This case also describes 2D experiments with scalar acoustics and electromagnetic waves. In the case of a membrane with fixed boundaries, one can describe the eigenfunction as a product of sine functions such that the wavevector k obeys the following equation $k^2=k_x^2+k_y^2$. As the membrane is a square cavity of dimension $L^2$, we can use two integers $(n,m)$ to describe the two components of the wavevector such that $k_x=n\pi/L$ and $k_y=m\pi/L$. Therefore we can describe the wavevector by $k^2=(n^2+m^2 ) \pi^2/L^2$. Since the dispersion relation is a linear function $\omega=\alpha k$, we obtain the following relation for the angular frequencies $\omega_{(n,m)}=\alpha\pi/L \sqrt{n^2+m^2}$. It is clear that the presence of the square root is going to give rise to frequency steps which are not commensurable. Consequently, the revival effect cannot occur in a square cavity with linear dispersion relation. We numerically confirm this in figure \ref{figure2}. The black thin line represents $C(t)$ for an elastic pulse propagating in a thin membrane. $C(t)$ for the membrane case presents very early fluctuations which correspond to the cavity classical time, $T_{\text{cl}}$ =70 $\mu $s. However, it does not present any significant fluctuation at longer time. This confirms the important  role of the dispersion for the revival effect. 

Finally, the influence of the boundary conditions in our experiment exemplified the essential role of the quadratic nature of the dispersion in the observation of quantum revival. We point out that in Figs.~\ref{figure1}a-c, the fluctuations of level spacing are of the same order, typically 0.5 kHz (vertical scale is different), but much larger than in Fig.~\ref{figure1}d. This is a direct consequence of the loss of commensurability between eigenfrequencies for the case of rigid boundaries. By choosing the source location on a high symmetry point, the smallest level spacing is artificially increased, as well as the separation between its multiples. Therefore the relative fluctuations with respect to the average mode spacing is the lowest in the case of a source centered (Fig.~\ref{figure1}a) while it is the largest for an arbitrary position of the source (Fig.~\ref{figure1}c). The choice of the source position is therefore an alternative to compensate for the non-quadratic nature of the dispersion relation and to favor the observation of quantum revival.
\\
\\
In conclusion, we have investigated the dynamics of an elastic wave packet in a thin plate. Flexural waves in thin plate fulfill the conditions needed to observe quantum revival. We demonstrate experimentally and confirm numerically the existence of fractional and full revival events. We show how the initial wave packet position may influence the phenomenon and take advantage of it to shorten the revival time and improve its observation, in a situation where the dispersion relation is not strictly quadratic (rigid boundaries). This spontaneous reconstruction of the initial wave packet depends only on the dimensions and geometry of the cavity and does not require any external active apparatus.
\\
\\
We would like to thank Roger Maynard who kindly initiated, inspired and fostered this work. M.D. acknowledges Ph.D. funding from the Direction G\'en\'erale de l'Armement (DGA). P.S. is thankful to the Agence Nationale de la Recherche support under grant ANR PLATON (No. 12-BS09-003-01), the LABEX WIFI (Laboratory of Excellence within the French Program Investments for the Future) under reference ANR-10-IDEX-0001-02 PSL* and the Groupement de Recherche 3219 MesoImage. This work has been partly supported by the Israel Science Foundation (Grant No. 1781/15 and
2074/15) and the United States-Israel Binational Science Foundation NSF/BSF grant No. 2015694.

\end{document}